\documentclass[a4paper,11pt]{article}
\usepackage{jcappub} 

\usepackage{orcidlink}

\newcommand{\Lm}{\mathcal{L}_m}

\title{\boldmath Density perturbations in nonminimally coupled gravity: symptoms of Lagrangian density ambiguity}

\author[a,b]{Miguel Barroso Varela \orcidlink{0009-0006-9844-7661}}
\affiliation[a]{Departamento de Física e Astronomia, Faculdade de Ciências, Universidade do Porto, Rua do Campo Alegre s/n, 4169-007 Porto, Portugal}
\author[a,b]{Orfeu Bertolami \orcidlink{0000-0002-7672-0560}}%
\affiliation[b]{Centro de Física das Universidades do Minho e do Porto, Rua do Campo Alegre s/n, 4169-007 Porto, Portugal}%

\emailAdd{up201907272@edu.fc.up.pt}
\emailAdd{orfeu.bertolami@fc.up.pt}

\abstract{The evolution of density perturbations is analysed in a modified theory of gravity with a nonminimal coupling between curvature and matter. We consider the broken degeneracy between the choices of matter Lagrangian for a perfect fluid, $\Lm=-\rho$ and $\Lm=p$, and determine the differences between their effects on the effective gravitational constant.  We review the result for $\Lm=-\rho$ in the quasistatic approximation and show how it can lead to unphysical singular behaviour for late-time dominating models. This divergent regime can be avoided when considering the fully non-quasistatic perturbative equations, although the higher-order nature of the nonminimally coupled theory and the requirement of a physically viable effective gravitational constant strongly constrains the magnitude of these modifications to the action. We find that both of these issues can be removed when considering $\Lm=p$ at late times due to the pressureless nature of non-relativistic matter and provide predictions for inverse power-law models. }

\begin{document}
\maketitle
\flushbottom


\section{Introduction}
The emergence of precision cosmology has provided evidence for some of the most recent paradigm shifts in modern physics. Abnormal galaxy rotation curves pushed for the need of dark matter \cite{DarkMatterReview2,DarkMatterReview}, while the observed accelerated expansion of the Universe brought the need for dark energy \cite{SupernovaSearchTeam:1998fmf}. Although many of the detected effects are accounted for by the standard model of cosmology ($\Lambda$CDM), this long-standing belief is now critically endangered by the ever-growing quality and quantity of cosmological data. Efforts from collaborations such as the Dark Energy Survey (DES) \cite{DES_main,DES_BeyondLCDM} and the Dark Energy Spectroscopic Instrument (DESI) \cite{DESI,DES_DESI_LateTimeDarkEnergy} have consistently tightened the constraints on the late-time properties of the Universe's expansion. Indeed, the latest data release from DESI indicates that the present accelerated expansion may be driven by evolving dark energy, which directly contradicts the cosmological constant included in $\Lambda$CDM at around $(1.7-3.3)\sigma$ depending on the supernovae data that is paired with BAO \cite{DESI:2025zgx}. Although this conclusion is not yet statistically significant enough to securely indicate a new discovery, the possibility of such a paradigmatic turn has further motivated research into alternatives to the current standard model. \par
Modifications of $\Lambda$CDM typically alter the matter content of the Universe, the underlying gravitational theory or a combination of both. In terms of the former, some examples include considering massive neutrinos \cite{DES_DESI_LateTimeDarkEnergy}, axions \cite{Rogers:2023ezo}, a quintessence field with a Higgs portal \cite{Bertolami:2007wb} and the generalised Chaplygin gas \cite{ChaplyginBilic,ChaplyginKamenshchik,Chaplygin}, among many others. More importantly for what concerns this work, modifications of General Relativity (GR) have been thoroughly researched in the literature as solutions to the present cosmological tensions and possible alternatives to explain dark energy and dark matter. These include $f(R)$ \cite{FRHubbleTension,FRHubbleTension2,FRHubbleTension3}, $f(T)$ \cite{FTHubbleTension,FTHubbleTension2} and $f(Q)$ gravity \cite{FQHubbleTension}, along with nonminimally coupled matter-curvature models \cite{NMCHubbleTension,BarrosoVarela:2024ozs}. 
\par
Although the main focus of the alternative models is to provide theoretical explanations of background quantities such as the expansion rate, most of these models also imply modified predictions for the dynamics at perturbation level, which affect cosmological aspects such as the formation of large scale structure. In fact, at this scale there is also uncertainty in the community, as measurements of the clustering of structures in the Universe by early and late methods give rise to the $\sigma_8$ tension \cite{sigma8_2,Sigma8}. This is yet another putative shortcoming of $\Lambda$CDM, again leaving space for modified theories of gravity to patch this observational incoherence. It is therefore as crucial for any theory's ambition to explain the background as it is for it to explain the perturbative dynamics of the Universe.
\par
In this work, we analyse the late-time evolution of cosmological density perturbations in a modified theory of gravity with a nonminimal coupling (NMC) between matter and curvature and their implications on the viability of late-time dominating versions of the theory, as well as what this imposes as constraints on the choice of perfect fluid Lagrangian density. {The nonminimally coupled $f(R)$ model was originally considered in Ref. \cite{ExtraForce} in the context of introducing an extra force that provided an alternative explanation to the effects of dark matter on galaxy rotation curves. However, it is important to note that non-linear couplings of matter with gravity had previously been analysed in the context of cosmic acceleration and the cosmological constant problem \cite{Dolgov:2003at,Dolgov:2003fw,Nojiri:2004bi,Nojiri:2004fw,Allemandi:2005qs}. Since the proposal of the NMC $f(R)$ theory, several extensions of these functional nonminimal couplings have emerged, such as $f(R,\Lm)$ \cite{Harko:2024sea}, $f(R,T)$ \cite{Harko:2011kv} and $f(R,T^{\mu\nu}T_{\mu\nu})$ \cite{Roshan:2016mbt} theories of gravity.} The present work builds on the research conducted in Refs. \cite{NMC_CosmologicalPerturbations_Nesseris,NMCCosmologicalPerturbations} in the context of the NMC $f(R)$ model \cite{ExtraForce}, where the dynamics of density perturbations were analysed on sub-Hubble scales by assuming a quasistatic approximation and choosing the perfect fluid Lagrangian $\Lm=-\rho$. The NMC $f(R)$ theory has also been extensively researched in the context of reproducing the late-time acceleration of the Universe \cite{NMCAcceleratedExpansion}, mimicking the dark matter profile observed in galaxy rotation curves \cite{NMCDarkMatter,NMCDarkMatter2}, fitting the latest low-redshift background data \cite{BarrosoVarela:2024ozs}, proposing a solution to the open Hubble tension \cite{NMCHubbleTension} and modifying the propagation of gravitational waves \cite{NMCGravWaves,NMCGWPolarisations}.
\par
Throughout this work, we consider a background Friedmann–Lemaître–Robertson–Walker (FLRW) metric in comoving coordinates 
\begin{equation}\label{BackgroundMetric}
    ds^2=a^2(\eta)(-d\eta^2+dx^2+dy^2+dz^2),
\end{equation}
where the conformal time $\eta$ is related to cosmic time through $dt=a(\eta)d\eta$. We denote derivatives with respect to $\eta$ with primes, derivatives with respect to $t$ with dots and define the comoving Hubble parameter $\mathcal{H}=aH=a'/a$, keeping $H=\dot a/a$. For simplicity, we perturb the background metric as $g_{\mu\nu}=\bar{g}_{\mu\nu}+a^2(\eta)h_{\mu\nu}=a^2(\eta)(\eta_{\mu\nu}+h_{\mu\nu})$, where we have denoted the unperturbed metric with a bar, although this notation will be dropped when the distinction is clear, as in the case where a quantity is already multiplied by a first-order perturbation. We focus our analysis on the scalar sector of perturbations, which we choose to fix in the Newtonian gauge such that the total line element reads
\begin{equation}\label{PerturbedLineElement}
\begin{aligned}
    ds^2=a^2(\eta)&\big[\left.-(1+2\Phi)d\eta^2
    +(1-2\Psi)\delta_{ij}dx^idx^j\big]\right.,
\end{aligned}
\end{equation}
which aligns with the typically chosen convention in the literature. This gauge choice has the added benefit of providing a clear analogy to the Newtonian limit of GR, where $\Psi$ and $\Phi$ act as the gravitational potentials. 
\par
{The sub-horizon expressions shown throughout this work are valid within a momentum range such that the sub-Hubble approximation can be assumed, i.e. $k\gg\mathcal{H}$, while ensuring we do not leave the linear regime. The former is satisfied by noting that the present comoving Hubble parameter is approximately $\mathcal{H}_0\sim2\times10^{-4}\ \text{Mpc}^{-1}$ and we can thus take this approximation to hold for $z\lesssim5$ due to the evolution $\mathcal{H}(z)$. This is satisfied in this investigation, as we work in the context of the late Universe, as long as we focus on scales with $k\gtrsim0.01\ \text{Mpc}^{-1}$, which is around 1-2 orders of magnitude greater than the comoving Hubble parameter. Non-linear effects can be disregarded for $k\lesssim0.2-0.3\ \text{Mpc}^{-1}$ \cite{Dodelson:2003ft}, which sets our sub-Hubble momentum range as $0.01\ \text{Mpc}^{-1}\lesssim k\lesssim0.3\ \text{Mpc}^{-1}$. Following the same logic, the results shown in our work are reliable over all redshift values $z\lesssim5$, as they include both the weak and strong NMC limits in their formulation.  }
\par
This paper is organised as follows. We present the nonminimally coupled model, its background and perturbed equations, along with the equations for both the considered Lagrangian choices in Section \ref{NMCSection}. We review the derivation of the density perturbation equation in the quasistatic approximation for $\Lm=-\rho$, followed by a discussion on its possibly problematic application to late-time modifications of GR in Section \ref{QuasistaticLRhoSection}. This is followed by a full analysis of density perturbations for the same Lagrangian with no quasistatic assumption, where we present the general method and the leading order results, in Section \ref{FullLRhoSection}. We explore the alternative choice of Lagrangian $\Lm=p$ and its implications on the quasistatic evolution of density perturbations at late times in Section \ref{QuasistaticLPressureSection}. We conclude in Section \ref{ConclusionSection} by discussing the obtained results and possible extensions of our work. We use the $(-,+,+,+)$ signature and set $c=8\pi G=1$, although $G$ may be restored in certain instances to simplify the presentation of results for comparative purposes.

\section{Nonminimally coupled model}\label{NMCSection}
\subsection{Action and field equations}
The nonminimally coupled $f(R)$ model is described by the action \cite{ExtraForce}
\begin{equation}
    S_{NMC}=\int dx^4 \sqrt{-g}\left[\frac{1}{2}f_1(R)+f_2(R)\mathcal{L}_m\right],
\end{equation}
where $f_{1,2}(R)$ are functions of the Ricci scalar $R$, $g$ is the determinant of the metric and $\Lm$ is matter Lagrangian density. We recover GR by setting $f_1=R$ and $f_2=1$ and minimally coupled $f(R)$ concerns the special case of $f_2=1$. A cosmological constant may be included in the theory via the curvature sector as $f_1=R-2\Lambda$. By varying the action with respect to the metric $g_{\mu\nu}$ we reach the modified field equations \cite{ExtraForce}
\begin{equation}\label{FieldEquations}
    FG_{\mu\nu}=f_2T_{\mu\nu}+\Delta_{\mu\nu}F+\frac{1}{2}g_{\mu\nu}(f_1-F R),
\end{equation}
where we have defined $\Delta_{\mu\nu}\equiv\nabla_\mu\nabla_\nu-g_{\mu\nu}\Box$, $F_i\equiv df_i/dR$ and $F=F_1+2F_2\mathcal{L}_m$. By taking the covariant divergence of both sides and considering the Bianchi identities $\nabla_\mu G^{\mu\nu}=0$ we obtain the modified conservation law \cite{ExtraForce}
\begin{equation}\label{NonConservationEq}
    \nabla_\mu T^{\mu\nu}=\frac{F_2}{f_2}\left(g^{\mu\nu}\mathcal{L}_m-T^{\mu\nu}\right)\nabla_\mu R,
\end{equation}
which is a unique consequence of the NMC sector of the theory, as seen by its direct dependence on $f_2$, and therefore reduces to the GR result when $f_2=1$. Both of these sets of equations show an important consequence of the introduction of a nonminimal coupling into the action - a direct dependence on $\Lm$, as seen explicitly in Eq. (\ref{NonConservationEq}) and through its presence in $F$ in Eq. (\ref{FieldEquations}). Unlike in GR, the gravitational dynamics depend directly on the matter Lagrangian and not uniquely on the related stress-energy tensor. \par

When considering the matter content of the Universe in a cosmological context, one typically assumes it is described by an isotropic and homogeneous perfect fluid with equation of state $p=w\rho$, where $p$ is the pressure and $\rho$ is the energy density. This can be described in terms of a diagonal energy-momentum tensor $T^\mu_\nu=\text{diag}(-\rho,p,p,p)$ which enters into the field equations in GR. To obtain this, one can consider the Lagrangian $\Lm=p$, which, upon variation with respect to the matter content and to the metric, yields the aforementioned diagonal tensor \cite{LagrangianForm}. 
However, as originally developed in Ref. \cite{LagrangianChoice2}, it was shown that the perfect fluid Lagrangian can be described by an action presented in terms of Lagrange multipliers that lead to particle number conservation, entropy exchange constraints and the flow of the fluid along the Lagrange coordinates. It was then shown that this leads to the on-shell Lagrangian $\Lm=p$, with the addition of surface integrals to the action leading to the equivalent choice $\Lm=-\rho$. Of course, this argument does not necessarily apply to theories with a nonminimal coupling between matter and curvature, such as the NMC $f(R)$ theory analysed in this work. As described in detail in Ref. \cite{LagrangianForm}, the original formulation from Ref. \cite{LagrangianChoice2} is mostly unaffected by the global multiplication by the NMC function $f_2(R)$, as 4 of the 6 equations of motion, derived from variation with respect to the different auxiliary variables, lead to the exact same result from GR, while the velocity representation and the temperature reflect the presence of the NMC. Substituting the equations in the action leads to the on-shell Lagrangian density $\Lm=p$. An analogous addition of surface integral then points to the same conclusion from Ref. \cite{LagrangianChoice2}, i.e. that $\Lm=-\rho$ can also be obtained as the on-shell Lagrangian for a perfect fluid.
As $\Lm$ is not explicitly present in the GR field and conservation equations, instead only appearing in the context of the associated energy-momentum tensor $T^\mu_\nu$, this choice has no effect on the gravitational dynamics. However, when considering the NMC model as above, we see that this degeneracy is broken by the explicit dependence on $\Lm$ in both the field and conservation equations. This was originally discussed in this context in Ref. \cite{LagrangianForm} and further considered in Refs. \cite{Faraoni:2009rk,NMCPerfectFluidDynamics,LagrangianChoice}. In this work, we shall analyse both of these options and investigate their impact on the dynamics of the NMC $f(R)$ model at the perturbative level through the evolution of density perturbations. 

\par
 For the remainder of this work we set $f_1(R)=R$ to isolate the effects of the nonminimal coupling function $f_2(R)$ on the theory's dynamics, although all effects of $f_1$ in the perturbative regime will be contained in $F$ and its derivatives, as there is no explicit dependence on $f_1$ apart from the background. 
 \subsection{Perturbed NMC equations}
{In line with our earlier definitions, we apply perturbations to the metric and the matter content of the theory by writing $g_{\mu\nu}=\bar g_{\mu\nu}+a^2(\eta)h_{\mu\nu}$, $\Lm=\bar{\mathcal{L}}_m+\delta\Lm$ and $T_{\mu\nu}=\bar T_{\mu\nu}+\delta T_{\mu\nu}$, such that the Ricci tensor is perturbed as $R_{\mu\nu}=\bar R_{\mu\nu}+\delta R_{\mu\nu}$, the Ricci scalar as $R=\bar R+\delta R$ and with the covariant derivative operators perturbed accordingly. Functions of the Ricci scalar can be perturbed by using the chain rule, i.e. by writing $f_i(R)=f_i(\bar R)+F_i(\bar R)\delta R$.} We thus obtain the perturbed field equations  \cite{NMCGravWaves}
    \begin{equation}\label{NMCLinearisedFieldEq}
\begin{aligned}
& \left(F_{1,R} \delta R+2  F_{2,R} \mathcal{L}_m \delta R+2  F_2 \delta \mathcal{L}_m\right) R_{\mu \nu}+\left(F_1+2  F_2 \mathcal{L}_m\right) \delta R_{\mu \nu} \\
& \quad-\frac{1}{2} g_{\mu \nu} F_1 \delta R-\frac{1}{2} a^2h_{\mu \nu} f_1-\left[\delta(\nabla_\mu \nabla_\nu)-a^2h_{\mu \nu} \square-g_{\mu\nu}\delta(\Box)\right]\left(F_1+2  F_2 \mathcal{L}_m\right) \\
& \quad-\left[\nabla_\mu \nabla_\nu-g_{\mu \nu} \square\right]\left(F_{1,R} \delta R+2  F_{2,R} \mathcal{L}_m \delta R+2  F_2 \delta \mathcal{L}_m\right) \\
& \quad=(1+ f_2) \delta T_{\mu \nu}+ F_2 T_{\mu \nu} \delta R,
\end{aligned}
\end{equation}
where we have defined $F_{i,R}=dF_i/dR$ to avoid confusion with the comoving time derivative $F_i'=F_{i,R}R'$. Again we see that the choice of Lagrangian density is non-trivial, as not only does it have an effect on the background quantity $\Lm$ but also on the perturbed matter content $\delta\Lm$. {The scalar curvature perturbation $\delta R$ can be calculated by explicit linear perturbation of the Ricci scalar quantity with a symbolic manipulation software or directly defined in terms of the metric fluctuations as} \cite{GWDecompositionBook}
\begin{equation}
    \delta R=\, -\bar R^{\mu\nu}a^2h_{\mu\nu}+\bar\nabla^\mu\bar\nabla^\nu(a^2 h_{\mu\nu})-\bar\Box(a^2 h) \, ,
\end{equation}
{where $h=\bar g^{\mu\nu}h_{\mu\nu}$ is the trace of $h_{\mu\nu}$ and we have maintained our convention of $g_{\mu\nu}=\bar g_{\mu\nu}+a^2(\eta)h_{\mu\nu}$. The first-order perturbation of the Ricci scalar is thus written in terms of the scalar metric perturbations as }
\begin{equation}\label{deltaR}
\begin{aligned}
    \delta R=-\frac{2}{a^2}\left[3\Psi''+6(\mathcal{H}'+\mathcal{H}^2)\Phi+\right.&\left.3\mathcal{H}(\Phi'+3\Psi')-k^2(\Phi-2\Psi)\right]
\end{aligned}
\end{equation}
where we have written spatial derivatives in Fourier space as $\nabla^2\Psi=-k^2\Psi$.\par
We write the Lagrangian-independent energy-momentum tensor and its scalar perturbations as \cite{GWDecompositionBook}
\begin{equation}
    \begin{aligned}
        & T^\eta_\eta=\bar T^\eta_\eta+\delta T^\eta_\eta=-(\rho+\delta \rho)=-(1+\delta)\rho \\
        &T^i_\eta= \delta T^i_\eta= -(\rho+ p)v^i=-(\rho+ p)\partial^i v=-\rho(1+c_s^2)\partial^i v\\
        &T^i_j= \bar T^i_j+\delta T^i_j=(p+\delta p) \delta^i_j=c_s^2(\rho+\delta\rho)\delta^i_j,
    \end{aligned}
\end{equation}
where we have defined $\delta=\delta\rho/\rho$ as the relative density contrast, $v$ as the scalar velocity perturbation and $c_s$ as the speed of sound in a perfect fluid, which relates its density and pressure via $p=c_s^2 \rho$. The fluctuations in the stress-energy tensor obey the perturbed conservation equation
\begin{equation}
\begin{aligned}
    \delta\left( \nabla_\mu T^\mu_\nu\right)=&\frac{F_2}{f_2}\left[(\delta^\mu_\nu \delta\mathcal{L}_m-\delta T^\mu_\nu)\partial_\mu R+(\delta^\mu_\nu\mathcal{L}_m-T^\mu_\nu)\partial_\mu (\delta R)\right]+\delta\left(\frac{F_2}{f_2}\right)(\delta^\mu_\nu\mathcal{L}_m-T^\mu_\nu)\partial_\mu R,
\end{aligned}
\end{equation}
where once again the choice of Lagrangian explicitly enters the dynamics at both background and perturbative levels.

\subsubsection{Perturbation dynamics in GR}
Under these assumptions and considering the sub-Hubble limit ($k\gg \mathcal{H}$), the perturbative analysis in GR gives Poisson equations for both of the equivalent scalar potentials \cite{GWDecompositionBook}
\begin{equation}
    \nabla^2\Phi=\nabla^2\Psi=-4\pi Ga^2\delta\rho,
\end{equation}
along with a second-order differential equation for the evolution of the relative density contrast
\begin{equation}
    \delta''+\mathcal{H}\delta'-4\pi G a^2\rho\delta=0.
\end{equation}
This shows how the evolution of the density perturbations can be considered to be scale-independent for sub-horizon scales. Typically their evolution over time is captured by the growth factor $D(z)$, defined as $D(z)=\delta(z)/\delta(0)$, such that $D(z=0)=1$ by convention. Additionally, we may define the growth rate  
\begin{equation}
    f_g=\frac{d\ln\delta}{d\ln a},
\end{equation}
which evidently is also scale-independent in GR. Considering a $\Lambda$CDM background, it has been found that to a high level of accuracy this quantity can be approximated by the fitting function $f_g=\left[\Omega_m(z)\right]^\gamma$, where $\Omega_m(z)=\Omega_{m,0}(1+z)^3/E(z)^2$ is the relative matter abundance and in flat $\Lambda$CDM we can write $E(z)=H(z)/H_0=\sqrt{\Omega_{r,0}(1+z)^4+\Omega_{m,0}(1+z)^3+\Omega_\Lambda}$. The growth index $\gamma$ has been constrained to $\gamma\approx0.55$ in $\Lambda$CDM \cite{Yin:2019rgm}. 

\subsection{Equations for \texorpdfstring{$\Lm=-\rho$}{L=-rho}}

We first consider the matter Lagrangian density $\mathcal{L}_m=-\rho$. The background equations provide two useful relations which can be used to simplify the analysis of the perturbations. Namely, combining the spatial and temporal parts of the modified field equations allows us to write
\begin{equation}\label{LoweringBackgroundEq}
    2F(\mathcal{H}^2-\mathcal{H}')+2\mathcal{H}F'-F''=a^2f_2\rho,
\end{equation}
which can be used to lower any second-order derivative $F''$ to a first-order derivative $F'$, thus ensuring we can combine terms in any intermediary steps. Additionally, the conservation equation is unchanged due to the choice of Lagrangian density, meaning that the evolution of the density of the fluid is still given by 
\begin{equation}
    \rho'+3\mathcal{H}(1+c_s^2)\rho=0.
\end{equation}
With these two conditions on the background evolution, we can ensure that the coefficients in the final differential equation for $\delta$ will only depend on $f_2$, $F$, $F'$ and $\rho$, along with the scale factor $a(\eta)$, its derivatives $\mathcal{H}^{(n)}$ and the wavenumber $k$. \par
The $(\eta\eta)$ component gives 
\begin{equation}\label{EtaEtaEq}
\begin{aligned}
    F\left[k^2(\Phi+\Psi)+\right.&\left.3\mathcal{H}(\Phi'+\Psi')+(6\mathcal{H}^2-3\mathcal{H}')\Phi+3\mathcal{H}'\Psi\right]
    +3F'(\Psi'-\mathcal{H}\Psi-3\mathcal{H}\Phi)=-f_2 a^2\rho\delta.
\end{aligned}
\end{equation}
By assuming the sub-Hubble approximation, valid for most scales of interest in late-time cosmology, this reduces to
\begin{equation}\label{PoissonEqDensityLagrangian}
    \Phi_{WL}\equiv\Phi+\Psi=-\frac{f_2}{F}\frac{a^2\rho}{k^2}\delta=-\Sigma\frac{a^2\rho}{k^2}\delta,
\end{equation}
where we have defined the weak lensing potential $\Phi_{WL}$ as well as the weak lensing parameter $\Sigma=f_2/F$, which reduces to 1 in GR, as expected. This has been analysed in the context of observationally fitted models in Ref. \cite{NMCGWPolarisations}, where for inverse power-law models (c.f. Eq. (\ref{PowerLawModelEq}) below) it was found that their prediction for $\Sigma$ is well within the bounds set by the DES and Planck collaborations \cite{WLPlanck2018,WLMeasurement}.   \par
The $(\eta i)$ component gives 
\begin{equation}\label{iEtaEq}
    F\left[\Psi'+\Phi'+\mathcal{H}(\Psi+\Phi)\right]+F'(2\Phi-\Psi)=-\rho(1+c_s^2)v
\end{equation}
and the $(ij)$ components with $i\neq j$ give
\begin{equation}\label{ijEqDensityLagrangian}
    \Psi-\Phi=\frac{\delta F}{F}=\frac{F_{1,R}\delta R}{F}-\frac{2F_2\delta+2F_{2,R}\delta R}{F}\rho,
\end{equation}
which breaks the equality between the two potentials seen in GR \cite{NMCCosmologicalPerturbations}. This is not a unique property of the NMC theory, as minimally coupled $f(R)$ theories show this feature by contributing to this equation through $F_{1,R}$. Similarly, the NMC function $f_2$ introduces a $\delta R$ term which behaves similarly to the minimally coupled theory with $F_1\rightarrow -2F_2\rho$. However, there is an additional contribution from $\delta\rho$ from the variation of the Lagrangian density which further highlights the reinforced coupling between the metric and density perturbations. 

\par
The $(ii)$ component involves the contribution of the pressure of the fluid, but is not shown here as it is not necessary in the calculations required for this Lagrangian density choice. In fact, the number of independent variables at play is such that this equation can be obtained from the remaining equations shown here and is therefore irrelevant when already considering all the others \cite{NMCGWPolarisations}. \par
The perturbed modified conservation equation provides two additional equations from the temporal and spatial components, respectively. Due to the choice of Lagrangian density $\mathcal{L}_m=-\rho$, the former is unaltered from GR, giving 
\begin{equation}\label{TemporalConservationEq}
    \frac{\delta'}{1+c_s^2}-3\Psi'-k^2v=0,
\end{equation}
while the latter is significantly changed by the nonminimal coupling, now giving
\begin{equation}\label{SpatialConservationEq}
    \frac{c_s^2}{1+c_s^2}\delta+\Phi+\left[\mathcal{H}(1-3c_s^2)+\xi_cR'\right]v+v'+\xi_c\delta R=0,
\end{equation}
where we have simplified $\delta\ln(f_2)=\frac{F_2}{f_2}\delta R=\xi_c\delta R$ by defining the NMC parameter $\xi_c\equiv F_2/f_2$. These equations can be combined to give a more complete description of the evolution of the density perturbations. In a late-time Universe dominated by matter ($c_s^2=0$) over radiation ($c_s^2=1/3$), we use Eq. (\ref{TemporalConservationEq}) to solve for $v$ and apply this to Eq. (\ref{SpatialConservationEq}) to get

\begin{equation}\label{SimplifiedSpatialConservationLawDensityLagrangian}
\delta''+\left[\mathcal{H}+\xi_cR'\right](\delta'-3\Psi')+k^2\Phi+k^2\xi_c\delta R-3\Psi''=0,
\end{equation}
which reduces to the GR result when we set $f_2(R)=1$, as expected. At this point, the unique NMC modifications are clear. These come in the form of a modification of the friction term $\mathcal{H}\rightarrow\mathcal{H}+\xi_cR'$ and an additional source term $\Phi\rightarrow\Phi+\xi_c\delta R$. The latter will have an important effect in this work's discussion, as it introduces higher-order terms in $k$ due to the interplay between curvature and density fluctuations in the NMC theory.

\subsection{Equations for \texorpdfstring{$\Lm=p$}{L=p}}
We now consider the alternative Lagrangian density choice $\Lm=p$. Similarly to the previous section, we find 2 independent background equations, with the only difference in the field equations being the form of $\Lm$ in $F$. However, the conservation equation is now modified to give
\begin{equation}
    \rho'+(1+c_s^2)(3\mathcal{H}+\frac{F_2}{f_2}R')\rho=0,
\end{equation}
which introduces an additional friction term due to the NMC function $f_2(R)$. \par
At perturbation level, the field equations are modified by the different Lagrangian choice.  The $(\eta\eta)$ component in the sub-Hubble limit now reads
\begin{equation}\label{EtaEtaEqLPressure}
    k^2\left[F(\Phi+\Psi)+2F_2\rho(1+c_s^2)(\Phi-2\Psi) \right]=-f_2a^2\rho\delta,
\end{equation}
which differs from $\Lm=-\rho$ by the addition of a $F_2\rho$ term, now appearing even when the Lagrangian depends only on the pressure, as it follows from the variation of $f_2T_\eta^\eta$, namely $F_2T^\eta_\eta\delta R$. This term is not present in the $\Lm=-\rho$ case due to the $(1+c_s^2)$ factor coming from $(\Lm+T^\eta_\eta)$, which is zero for that choice. \par
Apart from the change in the definition of $F$ through $\Lm$, the $(\eta i)$ component is precisely identical to what was found for $\Lm=-\rho$
\begin{equation}
    F\left[\Phi'+\Psi'+\mathcal{H}(\Phi+\Psi)\right]+F'(2\Phi-\Psi)=-(1+c_s^2)f_2a^2\rho v,
\end{equation}
while the $(ij)$ component for $i\neq j$ gives the same breaking of the degeneracy between the potentials
\begin{equation}
    \Psi-\Phi=\frac{\delta F}{F}=\frac{F_{1,R}\delta R}{F}+\frac{2F_2\delta+2F_{2,R}\delta R}{F}c_s^2\rho,
\end{equation}
only differing due to the switching of $-\rho\rightarrow p=c_s^2\rho$ for $F$. As we shall point out later, in a purely NMC model ($F_{1,R}=0$) and at late times, when the matter density is dominated by dust $(c_s^2=0)$, the equality of the two potentials is restored, even though there are still non-vanishing modifications to the remaining perturbative equations, as seen for example in Eq. (\ref{EtaEtaEqLPressure}). The same argument as before applies to the $(ii)$ component of the field equations, which can still be obtained from combinations of the others and is thus not considered here.\par
The most significant differences from the equations for $\Lm=-\rho$ arise from the perturbative conservation equations. In fact, the temporal component is now altered from GR, giving 
\begin{equation}
    \frac{\delta'}{1+c_s^2}-3\Psi'-k^2v+(\xi_c\delta R)'=0,
\end{equation}
where we used the fact that $R'\delta\xi_c=R'\frac{d\xi_c}{dR}\delta R=\xi_c'\delta R$ and have combined the $\xi_c\delta R$ terms using the product rule. The spatial component is now simpler, being given by 
\begin{equation}
    \frac{c_s^2}{1+c_s^2}\delta+\Phi+v\left[(1-3c_s^2)\mathcal{H}+c_s^2\xi_cR'\right]+v'=0,
\end{equation}
which shows how both components are now modified. However, by considering a dust-dominated late Universe ($c_s^2=0$), we see that the vanishing sound speed implies that the modification to the spatial component vanishes, analogously to what we had in the temporal component for $\Lm=-\rho$. In this late epoch, the equations can be combined to give 
\begin{equation}\label{SimplifiedConservationEqLPress}
    \delta''+\mathcal{H}\delta'-3\Psi''+(\xi_c\delta R)''+k^2\Phi-\mathcal{H}\left[3\Psi'-(\xi_c\delta R)'\right]=0,
\end{equation}
which again reduces to GR when we set $f_2(R)=1$. Note that the term responsible for the modified friction in the previous section is eliminated from the spatial component, since it is now proportional to $c_s^2$. 

\section{Quasistatic density perturbations for \texorpdfstring{$\Lm=-\rho$}{L=-rho}}\label{QuasistaticLRhoSection}
We start this section by reviewing the derivation in Ref. \cite{NMCCosmologicalPerturbations}. By considering Eq. (\ref{SimplifiedSpatialConservationLawDensityLagrangian}) in the sub-Hubble regime and using the quasistatic approximation, meaning that we neglect all time derivatives of the metric perturbations and keep only terms in the energy density fluctuations, we can write 
\begin{equation}
\delta''+\left[\mathcal{H}+\xi_c R'\right]\delta'+k^2\Phi+2\frac{k^4}{a^2}\xi_c(\Phi-2\Psi)=0,
\end{equation}
where we have written $\delta R$ under both approximations and used the NMC parameter $\xi_c\equiv F_2/f_2$. To solve for the potentials $\Psi$ and $\Phi$ in terms of $\delta$, we use their relationship from Eq. (\ref{ijEqDensityLagrangian}), which under the sub-Hubble and quasistatic approximations gives $\Phi(\Psi,\delta,k)$ and vice-versa for $\Psi$. This can be plugged into the Poisson Eq. (\ref{PoissonEqDensityLagrangian}) to give individual Poisson equations for $\Phi$ and $\Psi$. Plugging these into the equation above, we can get a second-order homogeneous differential equation for $\delta$
\begin{equation}
    \delta''+(\mathcal{H}+\xi_c R')\delta'-4\pi G_{\text{eff}}a^2\rho\delta=0,
\end{equation}
where the effective gravitational constant can be written as 
\begin{equation}
    \frac{G_{\text{eff}}}{G}=\Sigma\frac{1+4\frac{k^2}{a^2F}\left[F_{,R}-F\xi_c\left(1+3\frac{k^2}{a^2}\xi_c\right)\right]}{1+3\frac{k^2}{a^2F}F_{,R}},
\end{equation}
which incorporates the weak lensing parameter $\Sigma$ defined previously. This expression extends the minimally coupled $f(R)$ result, recovered when $\xi_c=0$ and thus $F=F_1$ and $\Sigma=1/F_1$. The NMC theory once again behaves similarly to its minimal counterpart by identifying $F_1\rightarrow -2F_2\rho$. However, there is still some additional dependence on $\xi_c$, which not only modifies the $k^2$ term in the numerator of $G_{\text{eff}}$ but also introduces a higher-order $\xi_c^2k^4$ term that follows from the $k^2\delta\ln(f_2)$ modification in Eq. (\ref{SimplifiedSpatialConservationLawDensityLagrangian}). \par
One natural concern when analysing the effective gravitational constant is the possibility of divergent behaviour when the denominator approaches zero. This can occur if $F_{,R}$ is negative and of large enough magnitude to have $k^2F_{,R}\sim\mathcal{O}(1)$. The sign of $F_{,R}$ is associated with the Dolgov-Kawasaki (DK) instability condition \cite{Dolgov:2003px}, which follows from associating an effective mass to the additional degree of freedom in $f(R)$ theories and imposing that it satisfies the condition $m_{f(R)}^2\propto F_{,R}>0$ \cite{NMCEnergyConditions}. For purely minimal theories, this simply implies $F_{1,R}>0$, while for pure NMC this requires $F_{2,R}\Lm>0$, which depends on the choice of $f_2$ together with the perfect fluid Lagrangian $\Lm$. This instability in the NMC model was considered in Refs. \cite{NMCEnergyConditions,NMCViability} for general power-law forms of $f_{1,2}(R)$ with positive exponents of $R$ that modify the dynamics of the early-time (high curvature) Universe.  \par
When considering late-time NMC modifications of GR, the function $f_2$ is taken to decrease as $R$ increases, such that we recover GR at early times. A natural choice is then any inverse power-law in $R$, as considered in Refs. \cite{NMCAcceleratedExpansion,NMCHubbleTension,BarrosoVarela:2024ozs}. However, the fact that $f_2>0$ and this inverse power-law dependence imply $F_{2,R}>0$, which coupled with $\Lm=-\rho<0$ leads to $F_{,R}<0$. In fact, as the scale of the NMC modifications was chosen such that the model reproduces the accelerated expansion of the Universe without a cosmological constant, the size of the $F_{,R}$ term in the denominator of $G_{\text{eff}}$ can grow to bring about divergent behaviour which is clearly unphysical. In the case of the previously mentioned models, this occurs for approximately $k\geq0.1\ \text{Mpc}^{-1}$. 
\par
This is a non-trivial result, although problematic behaviour was not discovered at background level, implying that for the same models the theory breaks down at the perturbative level. Even when temporarily disregarding this issue, the requirement of the effective gravitational constant not straying considerably from $G$ to ensure a physically viable evolution of the large-scale structure in the Universe imposes strong constraints on the magnitude of the NMC modifications due to the $k^4$ term in the numerator. \par
Another important consideration is the absence of ghost instabilities in a higher-order theory such as the one considered in this work. In minimally coupled $f(R)$ gravity, the absence of ghost instabilities is ensured by setting $df/dR>0$, separately from the condition $d^2f/dR^2>0$ required to avoid the emergence of Dolgov-Kawasaki instabilities \cite{DeFelice:2010aj}. In the context of nonminimally coupled $f(R)$ gravity, it is unclear what the corresponding condition would be, as the nonminimal coupling between matter and curvature makes attempts at fixing an analogous constraint inconclusive. This is particularly clear when attempting to directly derive the graviton propagator, which is unfeasible in the NMC theory, since the NMC implies that one cannot write the metric perturbation as given by the application of spin projectors to the energy-momentum tensor \cite{NMCViability}. 
As a ``zeroth-order" assumption, one could consider something similar to the condition for minimally coupled $f(R)$ theories, imposing instead that $F=F_1+2F_2\Lm>0$, which is indeed obeyed by all the models and Lagrangian choices considered in our work. Nonetheless, this theory, and particularly the specific model under consideration in this section, are intended to serve as an effective theory formulation of gravity at the cosmological level, and thus one cannot draw strict conclusions about the emergence of ghost instabilities.  Indeed, the power-law formulation of the $f_2(R)$ function is particularly useful in describing the theory as a power series that captures its behavior for a variety of curvature scales. Considering all of this, we should stress that the emergence of ghost degrees of freedom and the corresponding instabilities in the NMC $f(R)$ theory is an important piece in the big picture of investigations into this modified theory of gravity, and is worthy of its own detailed research work in the future.

\section{Full evolution of perturbations for \texorpdfstring{$\Lm=-\rho$}{Lm=-rho}}\label{FullLRhoSection}
Given the problematic nature of the denominator and the overall magnitude of the effective gravitational constant for cosmologically significant NMC models when imposing the quasistatic approximation, we now consider the equations obtained when the time derivatives of the metric perturbations are not neglected. Although we still expect the same $k^2$-dependence for large $k$ due to the modified conservation equations, it could be that the quasistatic approximation neglects contributions that keep the magnitude of $G_{\text{eff}}$ under control. In terms of the divergent behaviour, in Ref. \cite{delaCruz-Dombriz:2008ium} the full non-quasistatic analysis was conducted for minimal $f(R)$ theories and it was found that the leading term in the denominator of the effective gravitational constant is proportional to $(F_{1,R}k^2)^4$, which is always positive regardless of the sign of $F_{1,R}$. It could be that a similar term is present in the NMC theory, which would cancel the negative nature of the Lagrangian such that the singular behaviour of the perturbative analysis in this theory is fixed.

\subsection{Dimensionless parameters}
Writing the resulting equations in a dimensionless form can significantly simplify the analysis. This motivates the introduction of the dimensionless parameters $\kappa_n=\mathcal{H}^{(n)}/\mathcal{H}^{n+1}$ and $\mathcal{F}_n=F^{(n)}/(\mathcal{H}^nF)$, where $X^{(n)}$ denotes the $n$th order derivative, which track the evolution of the scale factor and the growth of the modified theory's effects, respectively. In GR, the $\kappa_n$ parameters for a matter-dominated Universe are constant and given by $\kappa_1=-1/2$, $\kappa_2=1/2$, $\kappa_3=-3/4$ and $\kappa_4=3/2$, while the $\mathcal{F}_n$ parameters are trivially 0, as $F=1$. We can also show that the derivatives of these parameters can be written as 
\begin{equation}
\begin{aligned}
    \kappa_n'&=\frac{\mathcal{H}^{(n+1)}}{\mathcal{H}^{n+1}}-(n+1)\frac{\mathcal{H}'\mathcal{H}^{(n+1)}}{\mathcal{H}^{n+2}}=\mathcal{H}\left[\kappa_{n+1} -(n+1)\kappa_n\kappa_1\right]
\end{aligned}
\end{equation}
and similarly 
\begin{equation}
    \mathcal{F}_n'=\mathcal{H}(\mathcal{F}_{n+1}-n\kappa_1\mathcal{F}_n-\mathcal{F}_i\mathcal{F}_1),
\end{equation}
which are useful if all intermediate calculations are carried out in this dimensionless form. The $R$-derivative of $F_2$, present in Eq. (\ref{ijEqDensityLagrangian}), can be written in terms of these parameters by exploiting the relationship $F_2'=R'F_{2,R}$ and thus writing 
\begin{equation}\label{RDerivativeEq}
    F_{2,R}=\frac{F_2'}{R'}=\frac{3(1-F)-F\mathcal{F}_1}{12\mathcal{H}^2(\kappa_2-2)\rho}a^2,
\end{equation}
where we have used the fact that the background Ricci scalar is given by $R=\frac{6}{a^2}\mathcal{H}^2(1+\kappa_1)$. We have written this expression in terms of $F=1-2F_2\rho$ in order to facilitate calculations, although in the final result it will be useful to separate $F$ into its GR and NMC parts in order to distinguish terms which vanish when $f_2(R)=1$ and those which are fully due to the standard Einstein-Hilbert action. The background Eq. (\ref{LoweringBackgroundEq}) can also be written in terms of these dimensionless variables as 
\begin{equation}
    2(1-\kappa_1)+2\mathcal{F}_1-\mathcal{F}_2=\frac{a^2f_2}{\mathcal{H}^2F}\rho,
\end{equation}
which provides a way of simplifying equations.\par
Alternatively, when dealing with a purely NMC theory, one may define the $\mathcal{F}_n$ parameters in terms of $F_2$ and its derivatives. This will prove useful when presenting the final results, as it better isolates the effect of $F_2$. With this in mind, we define $\tilde{\mathcal{F}}_n=F_2^{(n)}/(\mathcal{H}^n F_2)$ and $\tilde{\mathcal{F}}_0=F_2\rho$, such that we may rewrite the previous variables as 
\begin{equation}
    \mathcal{F}_1= \frac{2 \tilde{\mathcal{F}} _0 \left(3-\tilde{\mathcal{F}} _1\right)}{1-2\tilde{\mathcal{F}} _0}
\end{equation}
and 
\begin{equation}
   \mathcal{F} _2= \frac{2 \tilde{\mathcal{F}} _0 \left(3 \kappa _1+6 \tilde{\mathcal{F}} _1-\tilde{\mathcal{F}} _2-9\right)}{1-2 \tilde{\mathcal{F}} _0},
\end{equation}
which will be used in the presentation of results. 
\par

Another useful dimensionless parameter is $\epsilon=\mathcal{H}/k$, which serves as a quantity to compare the size of the perturbation's wavelength and the size of the cosmological horizon \cite{delaCruz-Dombriz:2008ium}. In the sub-Hubble limit, for which the characteristic wavelength of perturbations is considerably smaller than the horizon radius, we can treat this a small parameter ($\epsilon\ll1$) which can be used to write the result as a series expansion in which large powers of $\epsilon$ may be safely neglected. If one keeps this parameter throughout all calculations, it is important to keep in mind that it is a function of $\eta$, with its derivative being given by 
\begin{equation}
    \epsilon'=\frac{\mathcal{H}'}{k}=\epsilon\frac{\mathcal{H}'}{\mathcal{H}}=\mathcal{H}\kappa_1\epsilon.
\end{equation}

\subsection{Decoupling the equations}
We now simplify the coupled differential equations into a decoupled equation for the evolution of the density perturbations. For this, we use the method presented in Ref. \cite{delaCruz-Dombriz:2008ium}, where the full form of the equations was obtained for general minimally coupled $f(R)$ theories. In what follows, we briefly review this method, as it is still valid in the NMC model, albeit with significant added complexity. As we wish to focus on the late-time evolution of the Universe, we choose $c_s^2=0$ during all the following calculations. We start by removing the $v$-dependence from Eq. (\ref{iEtaEq}) by using the temporal component of the modified conservation law, Eq. (\ref{TemporalConservationEq}). We then combine the resulting equation with Eq. (\ref{EtaEtaEq}) in order to solve for expressions of $\Psi$ and $\Phi$ as functions of $\{\Psi',\Phi',\delta,\delta'\}$. We then differentiate these expressions with respect to $\eta$ and solve for $\Psi'$ and $\Phi'$ as functions of $\{\Psi'',\Phi'',\delta,\delta',\delta''\}$. We can use these expressions for $\Psi'$ and $\Phi'$ in Eqs. (\ref{iEtaEq}) and (\ref{EtaEtaEq}) to obtain $\Psi$ and $\Phi$ fully in terms of the same set of variables as for $\Phi'$ and $\Psi'$. \par
By solving for $\Psi''$ and $\Phi''$ in terms of $\delta$ and its derivatives, we can then write the metric perturbations as functions of these same variables, thus allowing for the determination of a decoupled ODE for the density perturbations. To do this, we start by replacing $F_{2,R}$ in Eq. (\ref{ijEqDensityLagrangian}) according to Eq. (\ref{RDerivativeEq}) and lowering the order of $\Psi''$ in $\delta R$ by using its dependence on $\{\Psi,\Phi,\Psi',\Phi',\delta,\delta',\delta''\}$ from the expressions determined thus far. We can then differentiate Eq. (\ref{ijEqDensityLagrangian}) and combine it with the spatial component of the modified conservation law with removed $v$-dependence, Eq. (\ref{SimplifiedSpatialConservationLawDensityLagrangian}), to find $\Psi''$ and $\Phi''$ as functions of $\delta$ and its derivatives of up to third order. At this point, we can use $\Phi(\delta,\delta',\delta'',\delta^{(3)})$ and $\Phi'(\delta,\delta',\delta'',\delta^{(3)})$ and the fact that the derivative of $\Phi$ must equal $\Phi'$ to find a fourth-order linear homogeneous differential equation for $\delta$. This final step can be taken equivalently with $\Psi$ and $\Psi'$ and leads to identical results, as expected. 

\subsection{Full density evolution equation}
We write the full fourth-order differential equation for $\delta$ as
\begin{equation}
A_4\delta^{(4)}+A_3\delta^{(3)}+A_2\delta''+A_1\delta'+A_0\delta=0,
\end{equation}
where the $A_i$ coefficients are functions of both conformal time and $k$, i.e. $A_i=A_i(\eta,k)$. The time dependence includes the contributions from the NMC modifications, as these follow from the background evolution, which is fully determined by the large-scale cosmological parameters and the NMC function $f_2(R)$. As discussed in Section \ref{NMCSection}, due to the higher-order nature of the NMC terms in the field equations, pure NMC terms are accompanied by larger powers of $k$ and thus lower powers of $\epsilon$. These will usually be dominant, unless the NMC function is small enough to counteract their magnitude superiority over the GR terms. In order to obtain a smooth GR limit, it is important to keep both the lowest order NMC and GR terms. All of this together leads to a non-quasistatic description of density perturbations in the sub-Hubble limit that can be evolved from the early to late Universe, fully fleshing out the effects of the modified theory on these physical quantities. \par
We split the coefficients into their GR and NMC components, labelling these as $\alpha_i$ and $\beta_i$ respectively. Unsurprisingly, $\alpha_3=\alpha_4=0$, while $\beta_3$ and $\beta_4$ are of higher order in $\epsilon$ and therefore may be approximately neglected in the sub-Hubble limit. By keeping only the lowest order terms in $\epsilon$ (or equivalently the highest in $k$), we get 
\begin{equation}
    \delta''+\frac{\tilde\alpha_1+\tilde\beta_1 k^{14}}{\tilde\alpha_2+\tilde\beta_2 k^{14}}\delta'+\frac{\tilde\alpha_0+\tilde\beta_0 k^{16}}{\tilde\alpha_2+\tilde\beta_2 k^{14}}\delta=0,
\end{equation}
where $\tilde\alpha_i$ and $\tilde\beta_i$ are the coefficients of the dominant terms for the GR and NMC sectors of the equation, which we have separated from their respective powers of $k$ for simplicity. We find that the effective gravitational constant, i.e. the quantity multiplying $\delta$ up to some factors, still exhibits scale-dependence, with the numerator evolving with a factor of $k^2$ relative to the denominator. However, these are now much higher-order in $k$, specifically being a factor of $k^{12}$ above what was found for the quasistatic result. This is analogous to the result in minimal $f(R)$ from Ref. \cite{delaCruz-Dombriz:2008ium}, where $G_{\text{eff}}$ gained a factor of $k^6$ on both the numerator and denominator, thus altering the usual $k^2$-dependence to a $k^8$-dependence. These parallels are expected, as the non-quasistatic approach builds on much of the same logic used in the quasistatic equations. In fact, the $k^2$ difference between numerator and denominator, as well as the difference of $k^6$/$k^{12}$ between the quasistatic and full approaches in the minimal/nonminimal theories, follow from the $k^2\xi_c\delta R$ term in Eq. (\ref{SimplifiedSpatialConservationLawDensityLagrangian}), which endows the NMC theory with unique higher-order behaviour in momentum space. The latter property of the NMC theory is carried over from the quasistatic result in the form of the modified $\delta'$ coefficient, also having a leading $k^{14}$ term instead of the $k^8$ seen in minimal $f(R)$ \cite{delaCruz-Dombriz:2008ium}.\par
The leading GR coefficients are given by 
\begin{equation}
    \begin{aligned}
    &\tilde\alpha_2=729\mathcal{H}^{14}(\kappa_2-2)^5(\kappa_1-1)^8\\
    &\tilde\alpha_1=729\mathcal{H}^{15}(\kappa_2-2)^5(\kappa_1-1)^8\\
    &\tilde\alpha_0=729\mathcal{H}^{16}(\kappa_2-2)^5(\kappa_1-1)^8(2\kappa_1-\kappa_2).
    \end{aligned}
\end{equation}
As the NMC terms are too complex to include here, we write each $\tilde\beta$ coefficient in the weak NMC limit. In this regime we find
\begin{equation}
    \begin{aligned}
    &\tilde\beta_2\approx-(\kappa_1-1)^2\tilde{\mathcal{F}}_0^6\tilde{\mathcal{F}}_1^2(-6+3\kappa_2+2(\kappa_1-1)\tilde{\mathcal{F}}_1)^4\\
    &\tilde\beta_1\approx-\mathcal{H}(\kappa_1-1)^2\tilde{\mathcal{F}}_0^6\tilde{\mathcal{F}}_1^2(-6+3\kappa_2+2(\kappa_1-1)\tilde{\mathcal{F}}_1)^4\\
    &\tilde\beta_0\approx-3(\kappa_1-1)(\kappa_2-2)\tilde{\mathcal{F}}_0^7\tilde{\mathcal{F}}_1(-6+3\kappa_2+2(\kappa_1-1)\tilde{\mathcal{F}}_1)^4  .
    \end{aligned}
\end{equation}
Note that $\tilde\beta_2\leq0$, while $\tilde\alpha_2\leq0$ if $\kappa_2\leq2$. As an exponentially expanding (de Sitter) Universe has $\kappa_2=2$ and we do not expect faster-than-exponential expansion in the NMC model \cite{NMCAcceleratedExpansion}, the condition $\kappa_2\leq2$ should be safely satisfied and therefore $\tilde\alpha_2\leq0$. This means that at leading order we do not observe the possibility of a null denominator and therefore avoid the singular behaviour found in the quasistatic result.
\par
As expected, the ratios of the $\tilde\alpha_i$ coefficients obey $\tilde\alpha_1/\tilde\alpha_2=\mathcal{H}$ and $\tilde\alpha_1/\tilde\alpha_2=\mathcal{H}^2(2\kappa_1-\kappa_2)$, such that when $f_2\rightarrow0$ we recover the GR equation
\begin{equation}
    \delta''+\mathcal{H}\delta'=-8\pi G(2\kappa_1-\kappa_2)\mathcal{H}^2\delta=4\pi Ga^2\rho\delta.
\end{equation} 
However, we should note that the same cannot be said about the $\tilde\beta_i$ coefficients, as for example $\tilde\beta_1/\tilde\beta_2\neq\mathcal{H}$. Interestingly, in the weak NMC limit, as shown above, we find that to lowest order 
\begin{equation}
\lim_{\tilde{\mathcal{F}_0}\to0}(\tilde\beta_1/\tilde\beta_2)=\mathcal{H},
\end{equation}
meaning that even in the extreme sub-Hubble limit we can still write the density perturbation differential equation as $\delta''+\mathcal{H}\delta'=4\pi G_{\text{eff}}a^2\rho\delta$ if we restrain ourselves to the weak NMC regime. In fact, if we consider a strong sub-Hubble limit where the ratio $k/\mathcal{H}$ compensates the weakness of the NMC components, we see can restrain ourselves to the $\tilde\beta$ coefficients, thus giving
\begin{equation}
\delta''+\mathcal{H}\delta'+\frac{3(\kappa_2-2)}{(\kappa_1-1)}\frac{\tilde{\mathcal{F}}_0}{\tilde{\mathcal{F}}_1} k^2 \delta=0   .
\end{equation}
As in the quasistatic analysis, we find that the effective gravitational constant has an overall $k^2$-dependence when considering a strong sub-Hubble regime. Remembering that $\tilde{\mathcal{F}}_1=F_2'/(F_2\mathcal{H})$ and taking the NMC functions to vary with the background as $F_2'\sim\mathcal{H}F_2$, one can estimate $\tilde{\mathcal{F}}_1\sim1$ and thus the size of the effective gravitational constant in the sub-Hubble regime is calculated as 
\begin{equation}
\begin{aligned}
     \frac{3(\kappa_2-2)}{(\kappa_1-1)}\tilde{\mathcal{F}}_0 k^2 \delta=\frac{6(\kappa_2-2)}{(\kappa_1-1)}F_2\frac{k^2}{a^2}\frac{a^2\rho\delta}{2}
     \Rightarrow G_{\text{eff}}=-6\frac{\kappa_2-2}{(\kappa_1^2-1)}F_2R\frac{k^2}{\mathcal{H}^2},
\end{aligned}
\end{equation}
where we used $R=6\mathcal{H}^2(\kappa_1+1)/a^2$. For a typical cosmological scale of $k=0.1 \ \text{Mpc}^{-1}$ we have at present $k/\mathcal{H}\sim400$, which means that to have an effective gravitational constant of the order of $G$, as constrained by the physically viable formation of large-scale structure, we need $F_2R$ to be $\mathcal{O}(10^{-5})$. For a power-law NMC function $f_2(R)$, this implies $(f_2-1)\sim F_2R\sim10^{-5}$, which constrains the NMC modifications to be considerably small, with no effective implications on the cosmological evolution at the background level. 

\section{Quasistatic density perturbations for \texorpdfstring{$\mathcal{L}_m=p$}{L=p}}\label{QuasistaticLPressureSection}

If instead of $\mathcal{L}_m=-\rho$ one considers the alternative $\mathcal{L}_m=p$ the equations are significantly different, particularly in the perturbed regime. However, if we are dealing with late-time effects of the theory then the dominant matter type is pressureless dust $(p_m=0)$, with a sub-dominant contribution from radiation that is effectively negligible in comparison to the non-relativistic matter density in the late Universe ($p_r=\rho_r/3\sim10^{-4}\rho_m$). With this in mind, it is clear that most contributions of the type $F_2\mathcal{L}_m$ will be small for this Lagrangian choice. In fact, in this regime we can write to a good approximation a mostly unmodified Friedmann equation
\begin{equation}
    \mathcal{H}^2\approx f_2\frac{a^2\rho}{3}+\mathcal{O}(F_2p),
\end{equation}
where the main change comes from the altered conservation equation for each matter type 
\begin{equation}
\begin{aligned}
    \rho'+(1+c_s^2)(3\mathcal{H}+\xi_cR')\rho=0\Rightarrow\rho\propto\left(a^3f_2\right)^{-(1+c_s^2)}\propto f_2^{-(1+c_s^2)}\hat\rho.
\end{aligned}
\end{equation}
Note that we have defined $\hat\rho$ as the density in GR, such that we separate the GR and NMC contributions to each density's evolution, which will simplify the analysis of the density perturbations. Importantly, this modified evolution is still present even when considering $F_2p\ll1$, and for the late-time dominating non-relativistic matter it implies an additional factor of $f_2^{-1}$ when compared to GR. However, this contribution is cancelled out in the Friedmann equation to give the typical evolution of the Hubble parameter $H^2=f_2\rho/3=\hat\rho/3\sim a^{-3}$. The inclusion of a cosmological constant would come from a modification of $f_1=R-2\Lambda$, leading to no contribution from $f_2$ and therefore the recovery of an exact $\Lambda$CDM background. \par
The same cannot be said about the equations in the perturbative regime. Apart from the typical $F_2\mathcal{L}_m$ terms and their perturbations, which would both be negligible, there are terms like $F_2\delta T^\mu_\nu$ which can lead to contributions such as $F_2\delta\rho$, which are as relevant as in the $\mathcal{L}_m=-\rho$ analysis. One major consequence of this regime is that $F\approx1$ and $\delta F\ll 1$, leading to a restoration of the equivalence of the scalar metric perturbations $\Phi\approx\Psi$, which greatly simplifies the resulting equations. For example, the $(\eta\eta)$ component of the perturbed field equations in the sub-Hubble limit and assuming negligible pressure reads
\begin{equation}\label{PressurelessPoissonEq}
\begin{aligned}
    \Phi=\Psi&=-\frac{f_2}{(1-F_2\rho)}\frac{a^2\delta\rho}{2k^2}=-\frac{f_2}{(1-\xi_c\hat\rho)}\frac{a^2\delta\rho}{2k^2}\equiv-\tilde\Sigma\frac{a^2\delta\rho}{2k^2},
\end{aligned}
\end{equation}
where we have defined the weak lensing parameter in this theory as $\tilde\Sigma=f_2/(1-F_2\rho)$ in analogy with the distinct parameter $\Sigma=f_2/(1-2F_2\rho)$ from the theory with $\Lm=-\rho$. We see that even for negligible pressure, which led to an unmodified background, we have non-negligible modifications in the Poisson-like equation connecting the metric and density perturbations.  \par
By combining the conservation Eq. (\ref{SimplifiedConservationEqLPress}) with the Poisson Eq. (\ref{PressurelessPoissonEq}) and taking the usual sub-Hubble limit together with the quasistatic approximation, we obtain a second-order differential equation for $\delta=\delta\rho/\rho$
\begin{equation}\label{DensityEqLPressure}
    \delta''+\mathcal{H}\delta'-4\pi \tilde G_{\text{eff}}a^2\rho\delta=0,
\end{equation}
where the effective gravitational constant is now given by 
\begin{equation}
    \frac{\tilde G_{\text{eff}}}{G}=\tilde\Sigma \left[1+\frac{2}{a^2}\left(2\xi_c(\mathcal{H}'-\mathcal{H}^2)+3\mathcal{H}\xi_c'-\xi_c''\right)\right],
\end{equation}
which is scale-independent and incorporates the weak lensing parameter $\tilde\Sigma$ analogously to what is obtained in the $\Lm=-\rho$ theory. This quantity depends on the evolution of the background curvature, taken to be identical to $\Lambda$CDM, through the NMC parameter $\xi_c$ and its derivatives, as well as directly on the background matter density through the weak lensing parameter $\tilde\Sigma$. 

\subsection{Pressureless density perturbations in models with \texorpdfstring{$f_2(R)=1+(R_n/R)^n$}{f2(R)=1+(Rn/R)n}}
We now turn to analysing the evolution of density perturbations as governed by Eq. (\ref{DensityEqLPressure}) within an inverse power-law NMC model that converges to GR at high curvatures (early times) but modifies their late-time evolution in a way that can perhaps better fit the available observational data and constrain the model's parameters. We take the background evolution to be exactly $\Lambda$CDM, with parameters taken from direct late-time measurements by the Pantheon+ collaboration \cite{Brout:2022vxf} $(\Omega_m=0.334,\ \Omega_\Lambda=0.666,\ H_0=73.3\ \text{km/s/Mpc})$. We consider as an example models with 
\begin{equation}
    f_2(R)=1+\left(\frac{R_n}{R}\right)^n,
    \label{PowerLawModelEq}
\end{equation}
such that $R_n$ sets the scale of the NMC effects. As mentioned before, the same kind of model was used in Refs. \cite{NMCAcceleratedExpansion,NMCHubbleTension,BarrosoVarela:2024ozs} in the context of background cosmological dynamics for $\Lm=-\rho$. For the sake of comparison between the two Lagrangian choices, we shall consider similar parameter spaces as those that led to pathological behaviour in the effective gravitational constant in Section \ref{QuasistaticLRhoSection}, with $n=4,6,10$ taken as exemplary cases and $R_n\sim3\times10^{-7}\ \text{Mpc}^{-2}$ \cite{BarrosoVarela:2024ozs}. Although in this case the background is unaltered from $\Lambda$CDM, it is important to show that the same choice of $f_2(R)$ can lead to completely distinct perturbative dynamics depending on the chosen perfect fluid Lagrangian.\par
\begin{figure}[t!]
    \centering
    \includegraphics[width=0.49\linewidth]{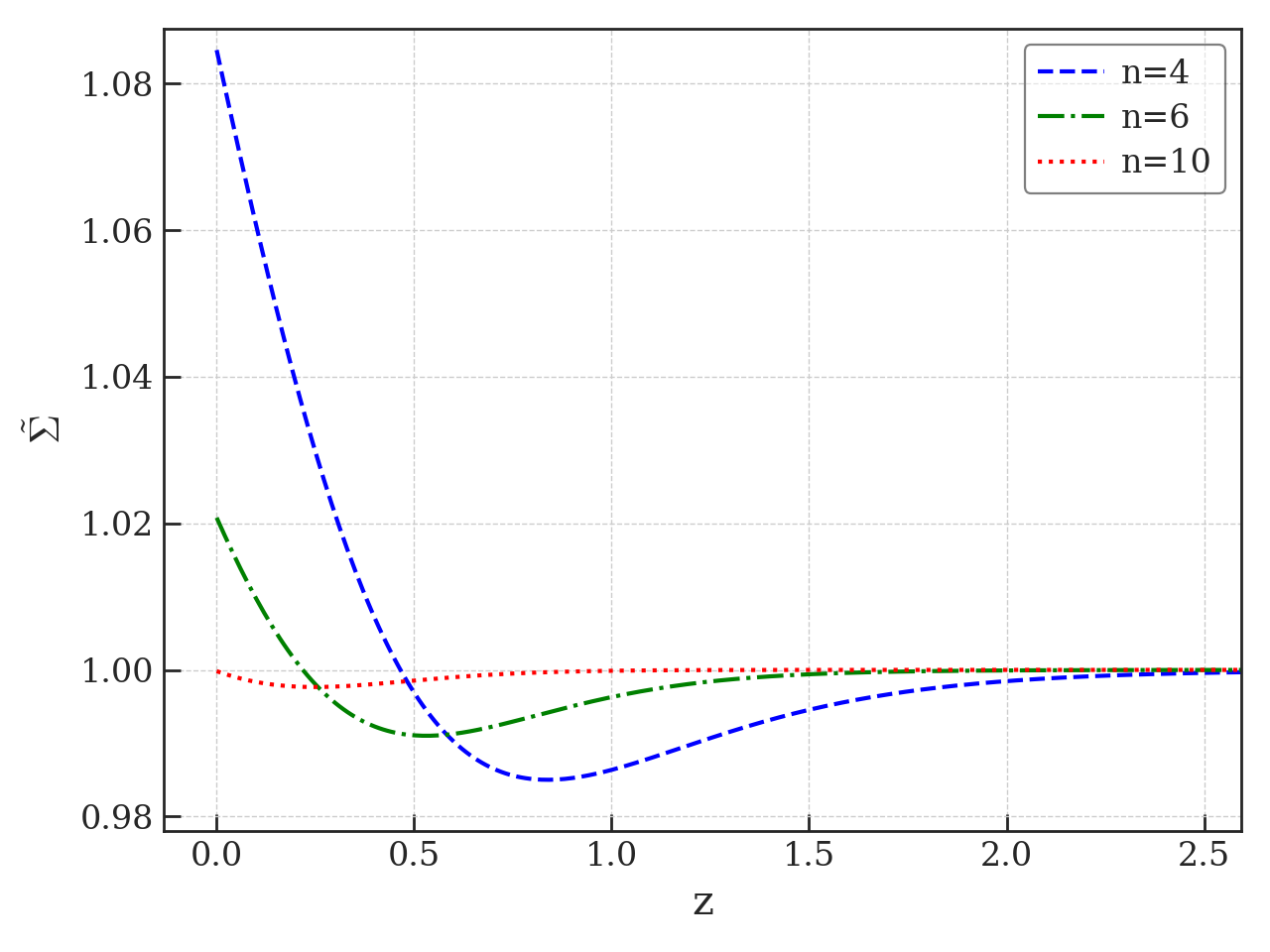}
    \includegraphics[width=0.49\linewidth]{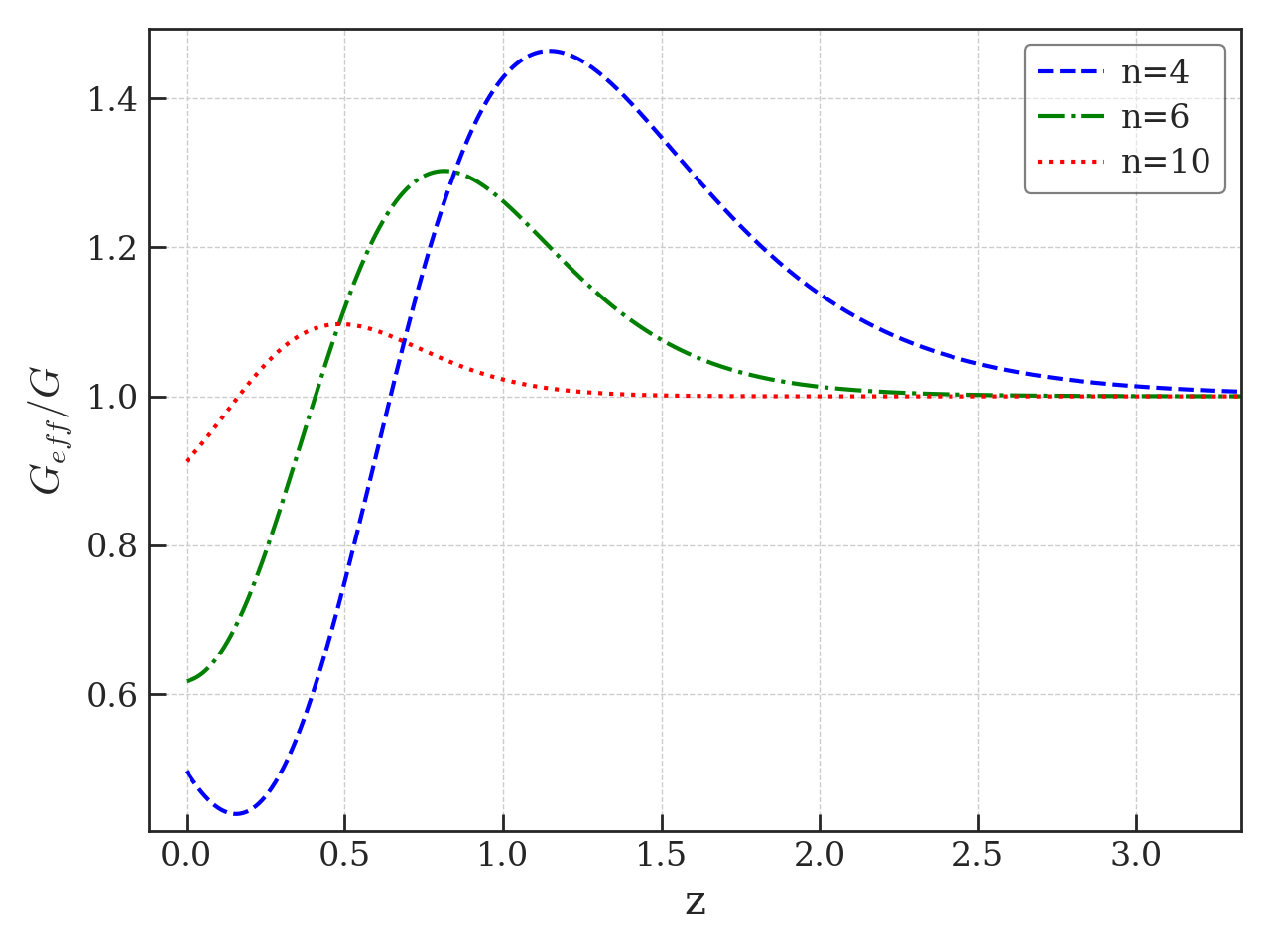}
    \caption{Weak lensing parameter $\tilde\Sigma$ (left) and effective gravitational constant $G_{\text{eff}}$ (right) in inverse power-law NMC models with $\Lm=p$. We have taken $R_n=3.3\times10^{-7} \ \text{Mpc}^{-2}$ for all models for comparative purposes.}
    \label{GEffFigure}
\end{figure}

The results for the evolution of the weak lensing parameter $\tilde\Sigma$ and the effective gravitational constant $G_{\text{eff}}$ in these models are shown in Figure \ref{GEffFigure}. We see that all considered models lead to a modification of the GR result for the weak lensing parameter, starting at $\Sigma_{GR}=1$ for redshifts larger than $z\sim2.5$ (for the chosen scale $R_n$) but then drifting to $\tilde\Sigma<1$ temporarily before crossing to $\tilde\Sigma>1$ at the present. This is all considerably similar to what was found  for $\Lm=-\rho$ in Ref. \cite{NMCGWPolarisations}, although the change of $2F_2\hat\rho$ to $F_2\rho=\xi_c\hat\rho<2F_2\hat\rho$ causes the denominator to be closer to 1 and therefore keeps the deviations from $\Sigma_{GR}$ smaller. Considering the results from Ref. \cite{NMCGravWaves} placed NMC theories within observational limits, it is natural that the tamer behaviour found here is also in line with the results from the Planck and DES collaborations \cite{WLPlanck2018,WLMeasurement}. The effective gravitational constant is kept within the order of magnitude of $G$, leading to at most deviations of $50\%$, although this can be softened or intensified depending on the chosen NMC scale $R_n$, as is the case with all exemplary effects considered in this section. Nevertheless, for the scales considered here we find that $G_{\text{eff}}$ starts by increasing from its GR value before peaking and decreasing to values below $G$. Physically this implies that matter experienced a stronger clustering around $z\sim1$ (varying depending on $n$ and $R_n$) but starts feeling a weaker attraction as the Universe begins accelerating near the present. 
\par
\begin{figure}[t!]
    \centering
    \includegraphics[width=0.49\linewidth]{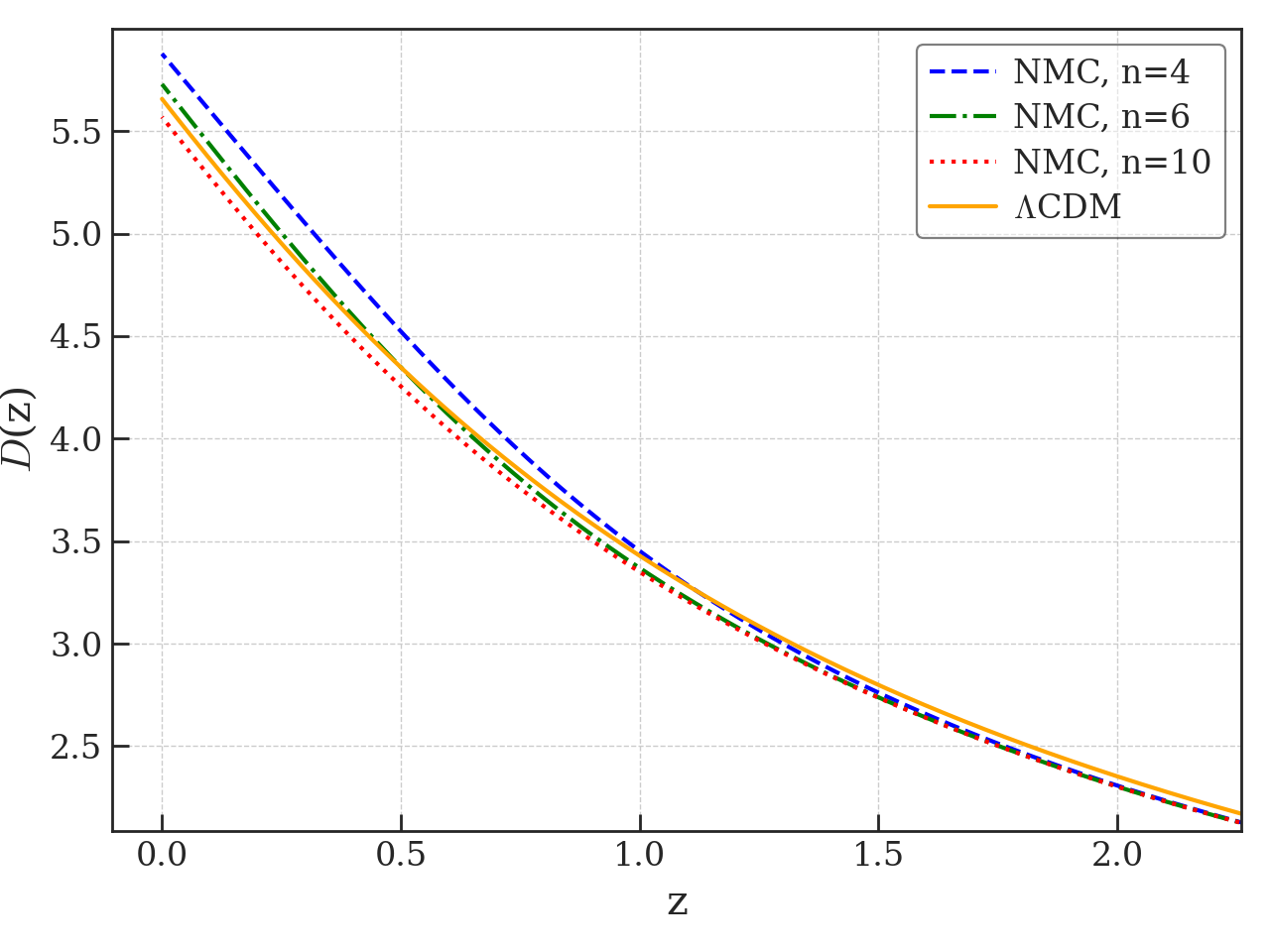}
    \includegraphics[width=0.49\linewidth]{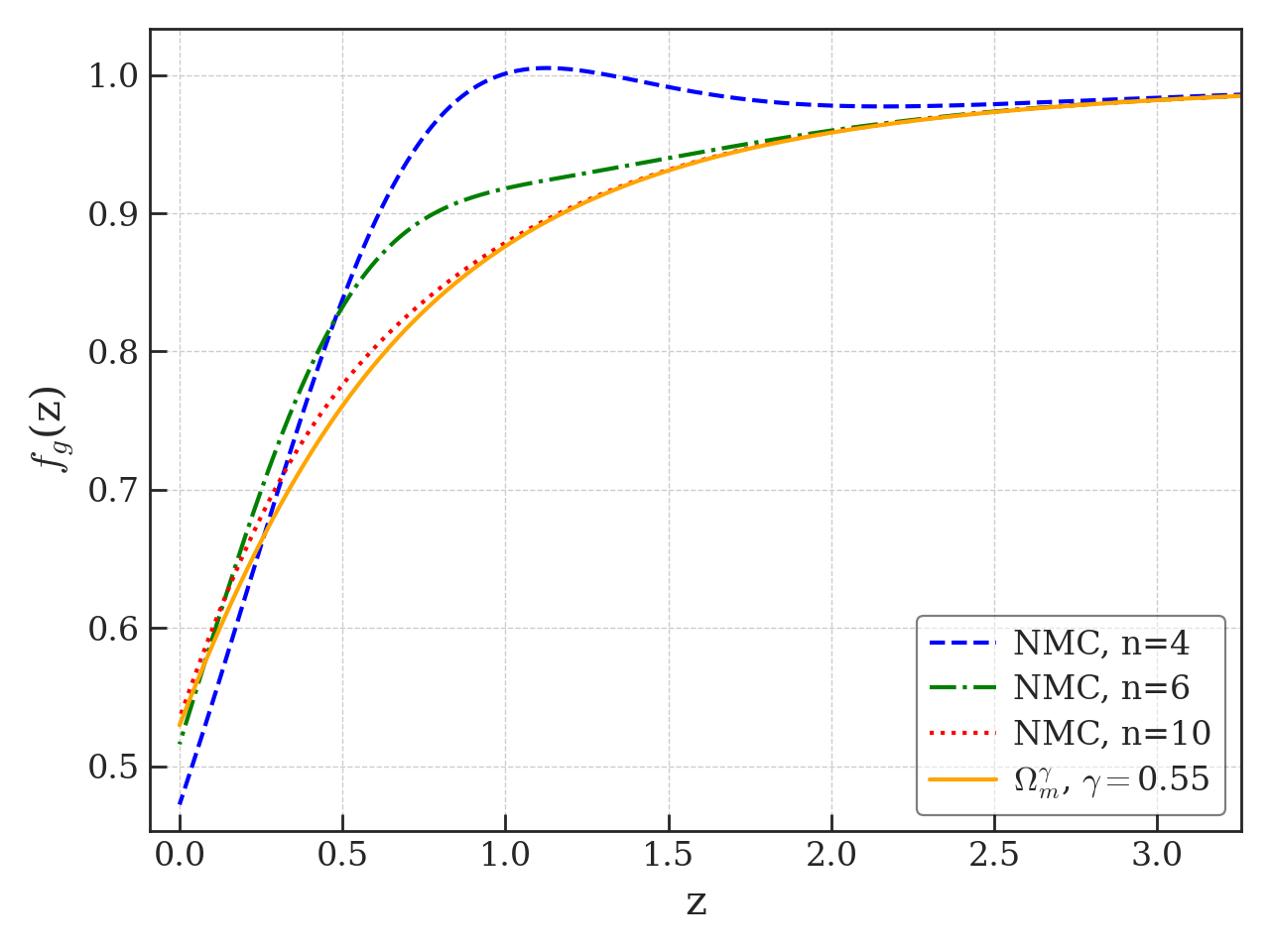}
    \caption{Normalised growth factor $D(z)$ (left) and growth rate $f_g(z)$ (right) in inverse power-law NMC models with $\Lm=p$. We have taken $R_n=3.3\times10^{-7} \ \text{Mpc}^{-2}$ for all models for comparative purposes.}
    \label{GrowthRateFigure}
\end{figure}

Inserting the calculated $G_{\text{eff}}$ in the density perturbation differential equation, we numerically evolve $\delta(z)$ and the associated growth rate $f_g(z)$. For this, we consider initial conditions taken from the $\Lambda$CDM model at redshifts of $z>8$, at which point the inverse power-law in $f_2(R)$ ensures a smooth return to GR. The growth rate can be estimated from the parametrisation $f_g\approx\Omega_m^{0.55}$ while the relative density fluctuations are estimated as $\delta(z_i)\sim a(z_i)\ll1$. We present the results for the normalised growth factor, here defined as $D(z)=\delta(z)/\delta(z_i)$ to clearly highlight any deviations from GR given a fixed initial condition $\delta(z_i)$, and $f_g$ in Figure \ref{GrowthRateFigure}. We observe that the growth rate converges to the GR parametrisation for redshifts larger than $z\sim2.5$, as observed in the weak lensing parameter, implying that the density fluctuations indeed evolve as $\delta\sim a$ before the NMC modifications become dominant. Due to the larger effective gravitational constant there is then an increase in the growth rate as the source term becomes stronger, causing a peak in $f_g$, deepening the gravitational potential wells and resulting in overdensities collapsing more efficiently. However, as we approach the present, this peak transitions into a sharp decrease in the growth rate, in some cases even becoming smaller than the present GR growth rate. \par
In fact, this behaviour is similar to that observed in Ref. \cite{Mirzatuny:2019dux}, where perturbations in the minimal $f(R)$ Hu-Sawicki model were considered and an accurate fitting function was proposed in place of the usual choice of $\Omega_m^{0.55}$. This proposed functional parametrisation is of the type 
\begin{equation}
    f_g^{(\text{fit)}}(z,k)=\Omega^{0.55}(z)\left[1+p(z,\kappa)\right],
\end{equation}
where $\kappa=k\lambda_C$. The parameter-dependent quantity $\lambda_C$ is the Compton wavelength associated with the modified theory's new degree of freedom, which is compared to the perturbation scale $k$ in order to determine the strength of the modifications. However, in the model analysed here we no longer observe scale-dependent perturbations to leading-order at late-times, which would in principle significantly simplify the search for such a fitting function. Once this is determined, one could accurately and efficiently generate predictions for a wide variety of observational probes, such as redshift space distortion, velocity correlation functions and galaxy-galaxy angular correlation, as described in Ref. \cite{Mirzatuny:2019dux}.

\section{Conclusions}\label{ConclusionSection}
In this work, we have analysed the late-time evolution of density perturbations in a nonminimally coupled theory of gravity. We considered two possible choices of the matter Lagrangian density for a perfect fluid, $\Lm=p$ and $\Lm=-\rho$, presenting the perturbed field and conservation equations for fluids governed by a general equation of state $p=c_s^2\rho$. We then reviewed the derivation of the $\Lm=-\rho$ density perturbation equations in the quasistatic approximation from Ref. \cite{NMCCosmologicalPerturbations} and presented arguments for why this raises pathological behaviour for pure NMC models with late-time cosmological implications. With these issues in mind, we extended the full non-quasistatic description of density perturbations in minimal $f(R)$ theories \cite{delaCruz-Dombriz:2008ium} to the NMC model, leading to higher powers of the wavenumber $k$ in both the effective gravitational constant and the friction term of the resulting differential equation for $\delta(z,k)$. Although at leading order we no longer find the singular behaviour observed in the quasistatic result, $G_{\text{eff}}/G$ is shown to significantly differ from 1 for cosmologically significant models, leading to irredeemable deviations from observations at the perturbative level. This is due to the effective constant evolving as $k^2$ for large $k$, which in the sub-Hubble limit imposes strong constraints on $f_2(R)$ and therefore holds back any hopes of dominant contributions at the background level, such as the reproduction of the accelerated expansion of the Universe. \par

Alternatively, we also focused on the option $\Lm=p$. In what concerns late-time modifications of GR, this choice effectively removes all NMC components from the modified Friedmann equation, following from the pressureless nature of the dominant non-relativistic matter, leaving only the modified evolution of the density. However, this dependence is cancelled in the Friedmann equation, restoring the typical $\Lambda$CDM background given the inclusion of a cosmological constant. The same is not true at the perturbative level, as although many terms are removed by the negligible pressure, the nonminimal coupling introduces some $\rho$-dependent terms. With this simplification, we showed the metric perturbations are identical, equivalently to GR, and derived the perturbative Poisson equation, which we used to determine the density fluctuation evolution in the quasistatic approximation. The effective gravitational constant is kept under control due to the sub-dominance of the pressure, allowing for non-pathological NMC modifications.  \par

{As we have restricted our discussion to the particular cases of $\Lm=-\rho$ and $\Lm=p$, it is natural to wonder what the effects of other Lagrangian forms would be. In fact, the addition of surface terms to the perfect fluid action described in Refs. \cite{LagrangianChoice2,LagrangianForm} allows for any perfect fluid Lagrangian of the form $\Lm=-\alpha\rho+(1-\alpha)p$, with $\alpha=0$ and $\alpha=1$ corresponding to $\Lm=p$ and $\Lm=-\rho$ respectively. We have not considered this general form, as any choice of $\alpha\neq0$ that is not fine-tuned to be suspiciously close to 0 ($\alpha\lesssim10^{-3}$) will behave as $\Lm\approx-\alpha\rho$ at late times due to the only contribution to the pressure being from the radiation content of the Universe (as $p=0$ for dust), which is significantly suppressed in comparison to the dust content during matter domination and necessarily remain that way onto the future. This choice would thus have the exact same pathologies as $\Lm=-\rho$, apart from some numerical differences in the exact results.}
\par
{The pathologies found for $\Lm=-\rho$ do not necessarily imply that $\Lm=p$ is the fundamentally correct choice from first principles. Even so, in Ref. \cite{Schutz:1970my}, Schutz presented the relativistic perfect fluid Lagrangian using a velocity potentials approach in an Eulerian picture, with the four-velocity being expressed in terms of six velocity potentials. This description leads precisely to the perfect fluid Lagrangian density $\Lm=p$. Unlike in Brown's approach \cite{LagrangianChoice2}, the density is a derived quantity from the pressure. In this example, one finds a derivation that obtains only $\Lm=p$ and not $\Lm=-\rho$. Another naive example is the application of this formulation to a perfect fluid described by a scalar field, as is the case with the inflaton or Higgs boson. In these scenarios, the pressure is associated with the kinetic and potential terms as $p=\dot\phi^2/2-V(\phi)$, while the density is associated with the total energy $\rho=\dot\phi^2/2+V(\phi)$. This means that identifying $\Lm=p$ leads to the typical description of the Lagrangian density for a scalar field, defined in terms of its kinetic energy minus its potential energy, while the same is not true for $\Lm=-\rho$. However, we should note that these are by no means decisive arguments, and the ambiguity regarding the correct choice of Lagrangian density for a perfect fluid remains an object of open debate in the non-minimal coupling research community \cite{Boehmer:2025afy,Fisher:2019ekh,Cipriano:2024jng,Harko:2024sea}. Finding such a decisive microscopic or variational argument that definitively rules out all but one option for $\Lm$ would be an immense contribution and that may hopefully be guided by more phenomenological arguments such as the ones presented in this work.}

\par
Our results point to so far overlooked issues in the perturbative dynamics of the NMC $f(R)$ model. The problematic nature of $G_{\text{eff}}$ in late-time dominating NMC models for $\Lm=-\rho$ does not necessarily point to their inviability, as it is plausible that one could construct a form of $f_2(R)$ that considerably affects the background evolution of the Universe with smaller NMC contributions, for example by keeping the cosmological constant in the action and instead focusing on introducing dynamical evolution to the dark energy equation of state, as observed by the DESI collaboration \cite{DESI,DESI:2025zgx}. This could also be achieved by analysing non-trivial forms of $f_1(R)\neq R$, as one would expect both minimal and nonminimal terms in a more complete effective theory of gravity. Not only could these have a relevant impact on the cosmological background, their effect on the perturbative framework presents the possibility of an interesting interplay between terms from $f_1(R)$ and $f_2(R)$ for either of the choices of Lagrangian density. Although this casts fundamental reservations on attempts at reproducing the accelerated expansion of the Universe via isolated effects from the NMC sector \cite{NMCAcceleratedExpansion,NMCHubbleTension}, it still leaves open the possibility of adding dynamical behaviour over a non-dynamical cosmological constant, introduced directly in $f_1=R-2\Lambda$ or in a minimal term that contains this behaviour, such as the Hu-Sawicki $f(R)$ model \cite{Hu:2007nk}.
Nevertheless, the issues raised in this work can serve as a way to considerably constrain the NMC $f(R)$ community's search for viable models for late-time cosmological evolution. This could also serve as a general argument when debating which perfect fluid Lagrangian should be chosen in $f(R)$ theories in which the Lagrangian degeneracy is broken. \par

A logical extension of this work is the application of the presented equations to the determination of the evolution of the growth factor $D(z,k)$ and growth rate $f_g(z,k)$ in these models, particularly their scale-dependence and how this differs from what is observed in minimal $f(R)$ theories. This can then be used to determine the matter clustering parameter $\sigma_8$, which can be compared to $f_g(z)$, $\sigma_8(z)$ and $f\sigma_8(z)$ data in order to constrain the parameters in any viable NMC function $f_2(R)$. Implications on the $\sigma_8$ tension \cite{Sigma8,sigma8_2} could then be determined and analysed similarly to what has been done for the Hubble tension \cite{NMCHubbleTension}.

\section*{Acknowledgments}
The authors would like to thank Álvaro de la Cruz-Dombriz and Andreas Mantziris for the useful discussions. The work of O.B. is partially supported by FCT (Fundação para a Ciência e Tecnologia, Portugal) through the project 2024.00252.CERN. The work of M.B.V. is supported by FCT through the grant 2024.00457.BD. Both authors are partially supported by UID/04650 - Centro de Física das Universidades do Minho e do Porto.

\bibliographystyle{JHEP}
\bibliography{References.bib}

\end{document}